\documentclass[fleqn,usenatbib]{mnras}

\usepackage[T1]{fontenc}
\usepackage{ae,aecompl}

\usepackage{graphicx}	
\usepackage{amsmath}	
\usepackage{amssymb}	

\title[]{Non-radial modes in classical Cepheids \\
What to look for in spectroscopy?}

\author[H. Netzel, K. Kolenberg]
{H. Netzel$^{1,2,3}$\thanks{E-mail: henia@netzel.pl},
K. Kolenberg$^{4,5,6}$\\
$^{1}$Nicolaus Copernicus Astronomical Centre, Polish Academy of Sciences, Bartycka 18, 00-716 Warszawa, Poland\\
$^{2}$Konkoly Observatory, Research Centre for Astronomy and Earth Sciences, E\"otv\"os Lor\'and Research Network (ELKH),\\ H-1121 Konkoly Thege Mikl\'os \'ut 15-17, Budapest, Hungary\\
$^{3}$MTA CSFK Lend\"ulet Near-Field Cosmology Research Group, H-1121 Konkoly Thege Mikl\'os \'ut 15-17, Budapest, Hungary\\
$^{4}$Institute of Astronomy, KU Leuven, Celestijnenlaan 200D, B-3001 Heverlee, Belgium\\
$^{5}$Physics Department, University of Antwerp, Groenenborgerlaan 171, B-2020 Antwerpen, Belgium\\
$^{6}$Department of Physics and Astronomy, Vrije Universiteit Brussel - VUB, Pleinlaan 2, B-1050 Brussel, Belgium\\
}

\date{Accepted XXX. Received YYY; in original form ZZZ}

\pubyear{2021}

\begin{document}
\label{firstpage}
\pagerange{\pageref{firstpage}--\pageref{lastpage}}
\maketitle

\begin{abstract}Recent photometric observations of first-overtone classical Cepheids and RR Lyrae stars have led to the discovery of additional frequencies showing a characteristic period ratio of 0.60-0.65 with the main pulsation mode. In a promising model proposed by \cite{dziembowski}, these signals are suggested to be due to the excitation of non-radial modes with degrees 7, 8 and 9 (Cepheids) or 8 and 9 (RR Lyrae). Such modes usually have low amplitudes in photometric data. Spectroscopic time series offer an unexplored and promising way forward. We simulated time series of synthetic line profiles for a representative first-overtone classical Cepheid model and added a low-amplitude non-radial mode. We studied sets of spectra with dense sampling and without noise, so-called 'perfect' cases, as well as more realistic samplings and signal-to-noise levels. Besides the first-overtone mode and the non-radial mode, also the harmonics of both modes and combination signals were often detected, but a sufficiently high sampling and signal-to-noise ratio prove essential. The amplitudes of the non-radial mode and its harmonic depend on the azimuthal order $m$. The inclination is also an important factor determining the detectability of the non-radial mode and/or its harmonic. We compared the results obtained for the predicted high degrees with those for lower-degree modes. Finally, we studied the sampling requirements for detecting the non-radial mode. Our findings can be used to plan a spectroscopic observing campaign tailored to uncover the nature of these mysterious modes.

\end{abstract}

\begin{keywords}
stars: variables: Cepheids -- stars: oscillations (including pulsations) -- techniques: spectroscopic
\end{keywords}



\section{Introduction}\label{sec.intro}

Classical Cepheids are radially pulsating stars, located in the classical instability strip in the Hertzsprung-Russell diagram. The majority of them pulsate in a single radial mode, either the fundamental mode (F) or the first overtone (1O). However, double-mode radially pulsating stars, and even triple-mode stars, are known among them \citep[see e.g.][]{cep_gc, cep_mc}. Recent photometric observations have shown a new kind of double-mode or triple-mode pulsations present in first-overtone Cepheids \citep[for a summary see][and references therein]{smolec_petersen}. In a fraction of classical Cepheids, and also in RR Lyrae stars, which pulsate in the first-overtone or first-overtone and fundamental mode (F+1O), an additional signal was detected with a period shorter than the period of the first-overtone mode. It forms a period ratio with the first-overtone mode in the range 0.60-0.65. This period ratio instantly suggests that the additional signal observed in these stars cannot correspond to any other radial mode \citep{dziembowski.smolec.2009}. Additional signals in these stars have low amplitudes, typically in the milimagnitude range. These additional signals form three characteristic, separate and parallel sequences in the Petersen diagram (i.e. a diagram of shorter to longer period ratio versus longer period) \citep[see e.g. Fig. 16 in][]{061cep_smc}. Interestingly, some of these stars have more than one additional mode, which corresponds to more than one sequence in the Petersen diagram. These additional frequencies were also found by \cite{AQLeo} in MOST satellite photomtery of the first discovered double-mode F+1O RR Lyrae star AQ Leo.  The authors detected two unidentified frequencies: a first one, $f_1$, shorter than the first-overtone frequency, and a second frequency, $f_2$, with $f_2=2f_1$, and $f_2$ forming this characteristic period ratio with the first-overtone mode in the range 0.60-0.65.

Large photometric surveys, like the Optical Gravitational Lensing Experiment \citep[OGLE, ][]{ogle}, led to the discovery of this additional signal forming a period ratio of 0.60-0.65 in dozens of Cepheids in the Large Magellanic Cloud \citep{soszynski_lmc_cep, moskalik.kolaczkowski.2009}, the Small Magellanic Cloud \citep{061cep_smc,soszynski_smc_cep} as well as in the Galactic disk \citep{pietrukowicz2013}. This signal was also detected in hundreds of RR Lyrae stars \citep[][and references therein]{netzel_census}), and shows similar characteristics (e.g. three sequences on the Petersen diagram) to the Cepheid sample. The TESS first-light results also provided one detection of a classical Cepheid with this additional mode \citep{tess}. Additionally, this signal was detected in an anomalous Cepheid for the first time by \cite{tess}. The excellent quality of space-based photometry from {\it Kepler} implies that three out of four first-overtone RR Lyrae stars observed in the original {\it Kepler} field show this signal \citep{068kepler}. All these results lead to the conclusion that this type of multi-mode pulsations must be common among first-overtone RR Lyrae stars and Cepheids. The origin of the signal, however, remains unknown.

\cite{dziembowski} proposed an explanation of the puzzling periodicity featuring non-radial modes of moderate angular degrees ($\ell$ 7 to 9) and periods longer than the period of the first overtone mode. Due to significant cancellation effects for modes of those relatively high degrees, it is easier to detect their harmonic, which forms this characteristic period ratio from 0.60 to 0.65. In Cepheids, these non-radial modes would have degrees $l=7,8,9$, each corresponding to one of three sequences on the Petersen diagram. In case of RR Lyrae stars the modes would have degrees $l=8,9$, which correspond to the bottom and top sequences on the Petersen diagram. The middle sequence is a combination of both modes.


To get more clarity on the physical origin of the additional modes in the first overtone classical pulsators, spectroscopy can be of help. Studying line profile variations (LPV) is a tool for mode identification, which gives information about the angular degree, $\ell$ and azimuthal order, $m$, of the mode. LPV analysis was already successfully used for different types of pulsating stars and various pulsation modes \citep[for an example of the application of LPV for $\delta$ Scuti stars see][]{lpv_example}.

In this paper, we make an attempt at answering the following question:
{\it If the additional modes are non-radial and of (relatively) high degree, can they be detected in spectroscopic data?} To this end, we carried out simulations for a plausible first-overtone Cepheid, producing spectroscopic line profile variations across several pulsation cycles for various mode parameters for the additional non-radial mode. In these generated spectra, we looked for the signatures of the non-radial mode.  We carried out this simulation for "ideal" noiseless data with optimal coverage, followed by a more realistic case study with "normal" time sampling for ground-based spectroscopy and signal-to-noise ratios. An additional question we will try to answer is:
{\it If we can detect a signal at the relevant frequencies in spectroscopic data, would we be able to constrain the mode parameters?} 

As was noted above, these additional signals are present in classical Cepheids and RR Lyrae stars, but we focus only on classical Cepheids in further investigation. Classical Cepheids are brighter than RR Lyrae stars and have longer pulsation periods. Hence, they are more promising targets for spectroscopic observations aimed at detecting low amplitude modes.

This paper is structured as follows: in Sec.~\ref{sec.method} we present the tools and methods that we used in the analysis. In Sec.~\ref{sec.Results} we present results and discuss them in Sec.~\ref{sec.discussion}. Finally, in Sec.~\ref{sec.summary} we sum up our findings.

\section{Simulations}\label{sec.method}

For simulating and analyzing spectra, we used FAMIAS \citep[Frequency Analysis and Mode Identification for AsteroSeismology, ][]{famias}. FAMIAS provides tools for line profile synthesis, frequency search in a set of spectra with Fourier analysis and non-linear least-squares fitting tools and mode identification. We used the first two capabilities to simulate synthetic line profiles and search for periodicities in resulting sets of spectra.

\subsection{Stellar parameters}

We generated synthetic line profiles for a first-overtone Cepheid with a period of 2.035 d from the Large Magellanic Cloud, OGLE-LMC-CEP-2532, studied by \cite{OGLE-LMC-CEP-2532}. However, we note that the frequency analysis of the available photometry for this star does not result in a detection of the additional signal of interest. As a member of a binary system, OGLE-LMC-CEP-2532 has relatively precisely determined physical parameters. We adopted the mass, temperature, metallicity and radius determined from previous studies \citep[see tab. 6 in][]{OGLE-LMC-CEP-2532}. The log$g$ value obtained for OGLE-LMC-CEP-2532 is 2.1. However, in FAMIAS, the lower limit for gravity is 2.5, but our simulations showed that in this gravity range the effect of the gravity on the amplitudes of the modes is negligible.  Therefore we set this lower limit for the simulations. Typical values of $v\sin i$ for first-overtone Cepheids are of the order of 20 km/s \citep{vsini_f1o_cep}. We used the relation $v_{rot} \sin i = -11.5 \log{P} + 19.8$ \citep{nardetto} and obtained 16.2 km/s. For the simulations, we adopted the value of 20 km/s. The inclination of OGLE-LMC-CEP-2532 is 85$^\circ$. In the next sections, we study the effect of the inclination for the non-radial mode's detectability. Still, for most synthetic line profiles, we adopted an inclination of 45$^\circ$, unless stated otherwise.

\subsection{Pulsation mode parameters}

We assumed two pulsations modes for each simulation: the first overtone and the additional non-radial mode of degree $\ell$=7, 8 or 9. We set the azimuthal order for values in a range ${\mathit m}\in(-\ell,\ell)$. These are considered the two main modes present in the star, and for this reason, we did not include any harmonics or combination frequencies. As we will see, these will naturally appear in the frequency spectra of the generated spectroscopic data.

\begin{figure}
\centering
\includegraphics[width=0.5\textwidth]{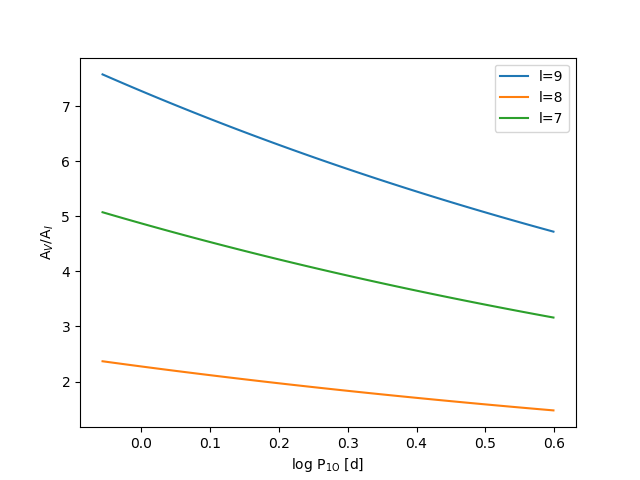}
\caption{Velocity amplitude and $I$ filter amplitude ratio as a function of first-overtone period for non-radial modes $l=7,8,9$.}
\label{fig.ampax}
\end{figure}

We adopted the velocity amplitude of the radial first overtone to be 28.2 km/s \citep[see Tab. 13 in][]{araucaria_binaries}.  
The frequency of the additional mode was adopted to fit the observed period ratio for given $\ell$. In order to get the velocity amplitude of the non-radial mode for a given $\ell$, we used Eq.~(8) from \cite{dziembowski2012}, with the Eddington approximation for limb darkening:

\begin{equation}
\frac{A_V}{A_I}=0.047\frac{R}{P_{\rm 1O}}\left[b_{E,\ell}\ell(\ell+1)\right]^{-1} [km/s/mmag],
\end{equation}

where $A_V$ is the velocity amplitude, $A_I$ is the amplitude in the I filter, which we set to 2 mmag, R is the photospheric radius, $P_{\rm 1O}$ - the first-overtone period, and $b_{E,\ell}$ - is the limb darkening in the  Eddington approximation for a given degree of the mode. In Fig.~\ref{fig.ampax} we plotted the ratio $\frac{A_V}{A_I}$ for a range of first-overtone periods and three non-radial modes.  To observe the same amplitude in the $I$ filter, the highest velocity amplitude is required for the mode of degree $\ell=9$ and the smallest for the mode of degree $\ell=8$.


In the case of the studied star based on OGLE-LMC-CEP-2532, the values of the velocity amplitudes for $\ell$=7,8 and 9 are 10.19 km/s, 4.75 km/s and 15.21 km/s, respectively. We checked whether the phase difference between the radial and the non-radial mode lead to significant changes. Based on several simulations with various phase differences, we concluded that it is not affecting the results. In the rest of the simulations, we set the radial mode phase to 0.0 and the non-radial mode phase to 0.15 (in $2\pi$ units).

\subsection{Other parameters}
We used the following default settings for the line profile parameters: a central wavelength at 5300 \r{A}, an equivalent width of 10 km/s and intrinsic width of 10 km/s. The adopted value for the intrinsic width is in agreement with values determined by the \cite{nardetto} ($\sigma_C$ in tab.~2), which is around 0.2-0.3 \r{A} for Cepheids. The values of the equivalent width can be calculated from values presented in Tab. 3 in \cite{nardetto}. 

For the noiseless "perfect" case, we simulated time series spanning 30 days with a step of 0.005 d (7.2 minutes), which results in 6001 spectra per simulation. The number of segments on the stellar surface for creating a line profile was set for 1000 in all cases. To simulate more realistic observations, we included gaps between the data and considered a much lower number of spectra. Also, we changed the signal-to-noise ratio by adding noise to the spectra.

\begin{figure*}
\begin{minipage}{180mm}
\centering
\includegraphics[width=\textwidth]{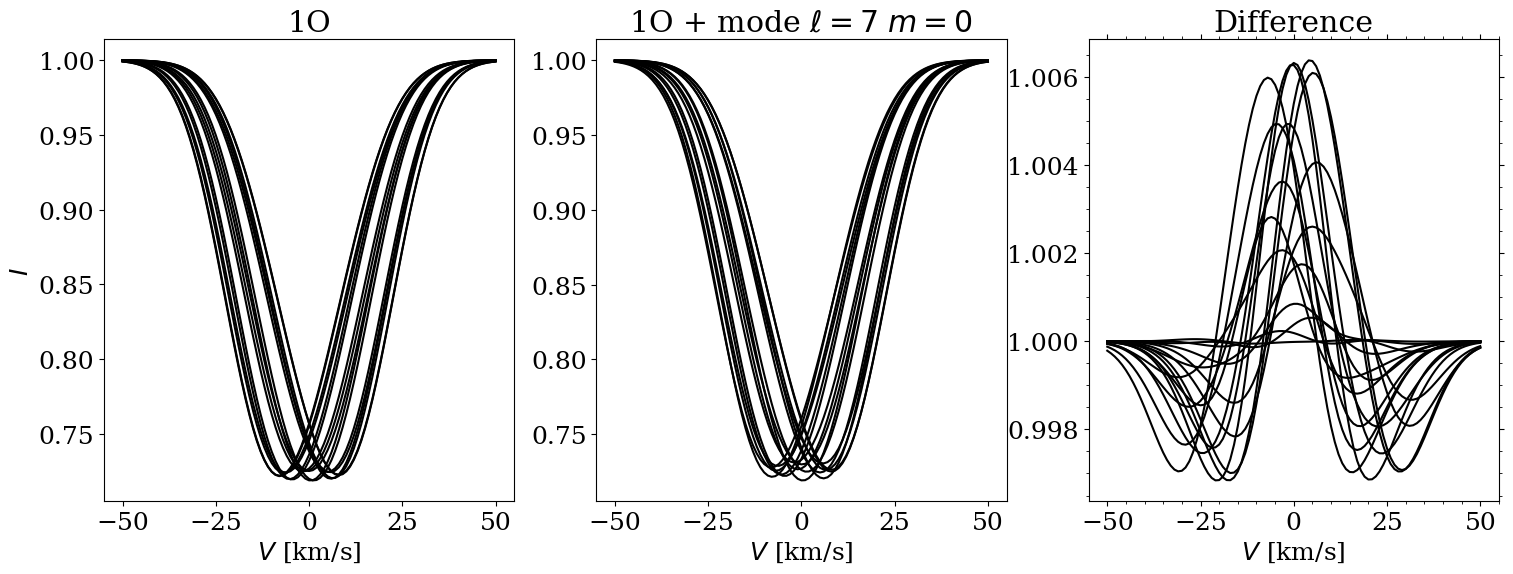}
\caption{A subset of simulations from our Line Profile Variations (LPV) sample. Adding an axisymmetric non-radial mode of degree $\ell = 7$ (middle panel) to the main pulsation with the first overtone mode (left panel) introduces small changes in the LPV (right panel). In this paper we investigate whether we can identify the nature of the additional (non-radial) mode from these differences.  }
\label{fig.lpv}
\end{minipage}
\end{figure*}

\begin{figure*}
\begin{minipage}{180mm}
\centering
\includegraphics[width=\textwidth]{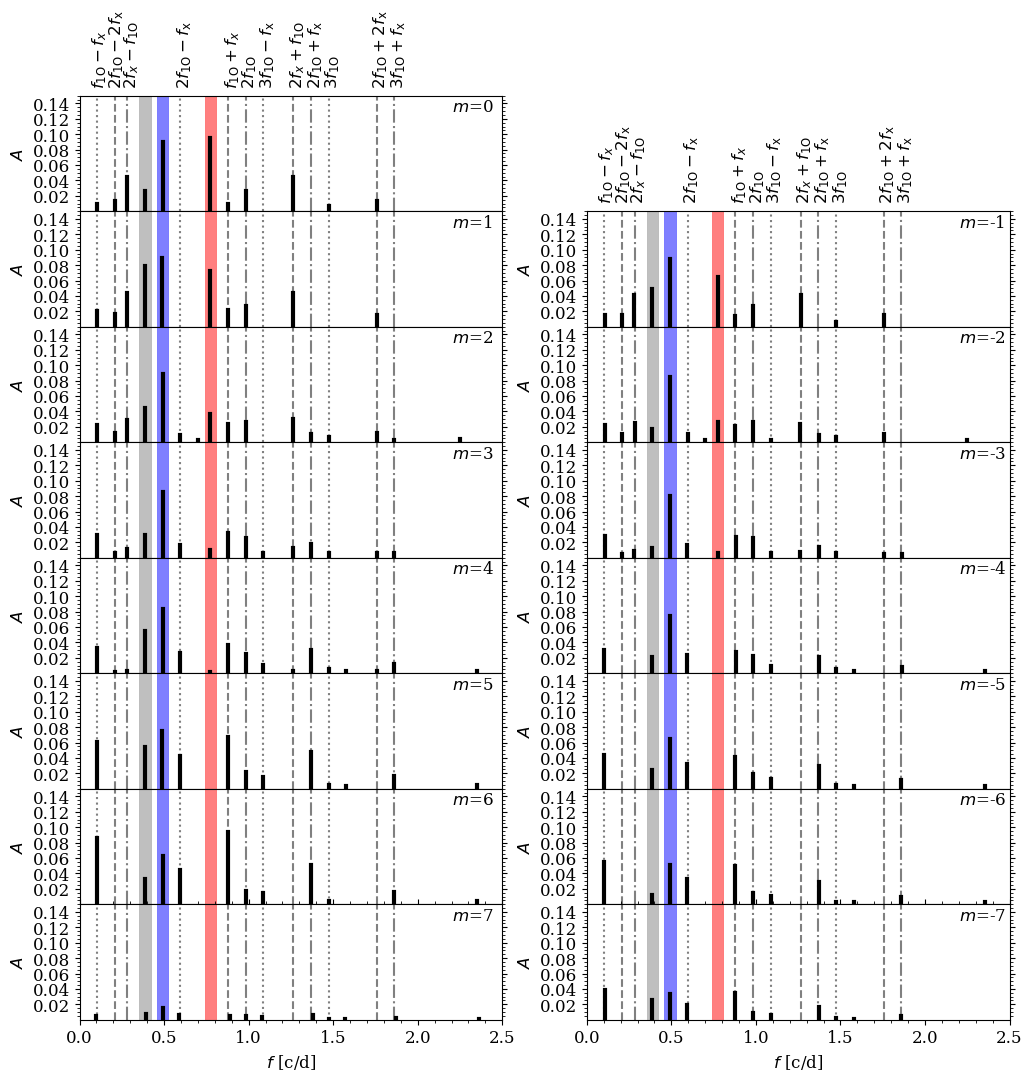}
\caption{Frequencies detected for the simulations using the non-radial mode of degree $\ell=7$. The left column shows results for positive $m$, and the right column for negative $m$ values. Blue, grey and red regions the show positions, within the frequency resolution of our simulations, of the first overtone mode, the additional mode and its harmonic.  Dashed lines mark harmonics of the signals and combination frequencies.}
\label{fig.l7_fourier}
\end{minipage}
\end{figure*}

\begin{figure}
\centering
\includegraphics[width=0.5\textwidth]{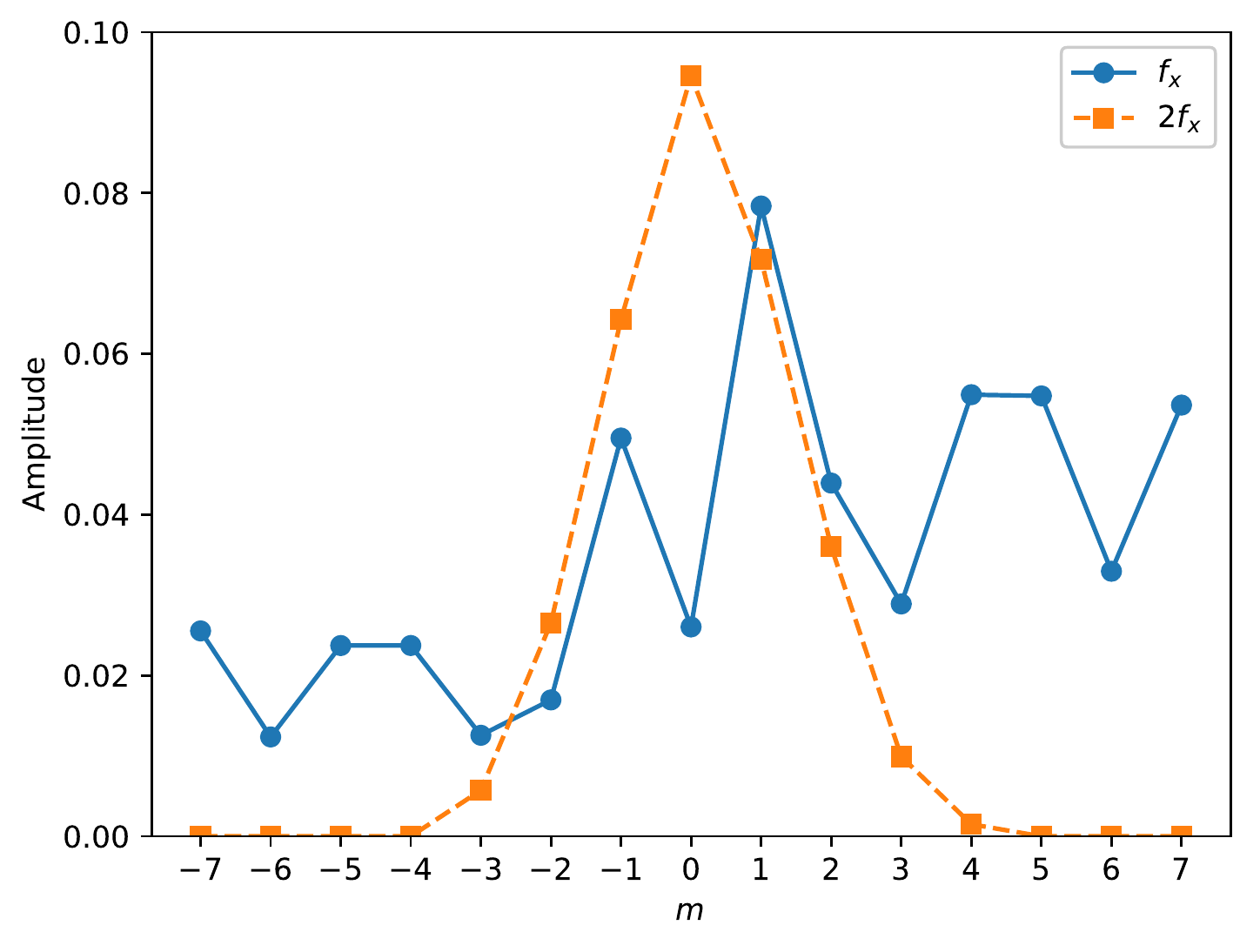}
\caption{Amplitudes of the non-radial mode with $\ell=7$ and its harmonic as a function of $m$.}
\label{fig.l7_amplitudes}
\end{figure}

\begin{figure}
\centering
\includegraphics[width=0.5\textwidth]{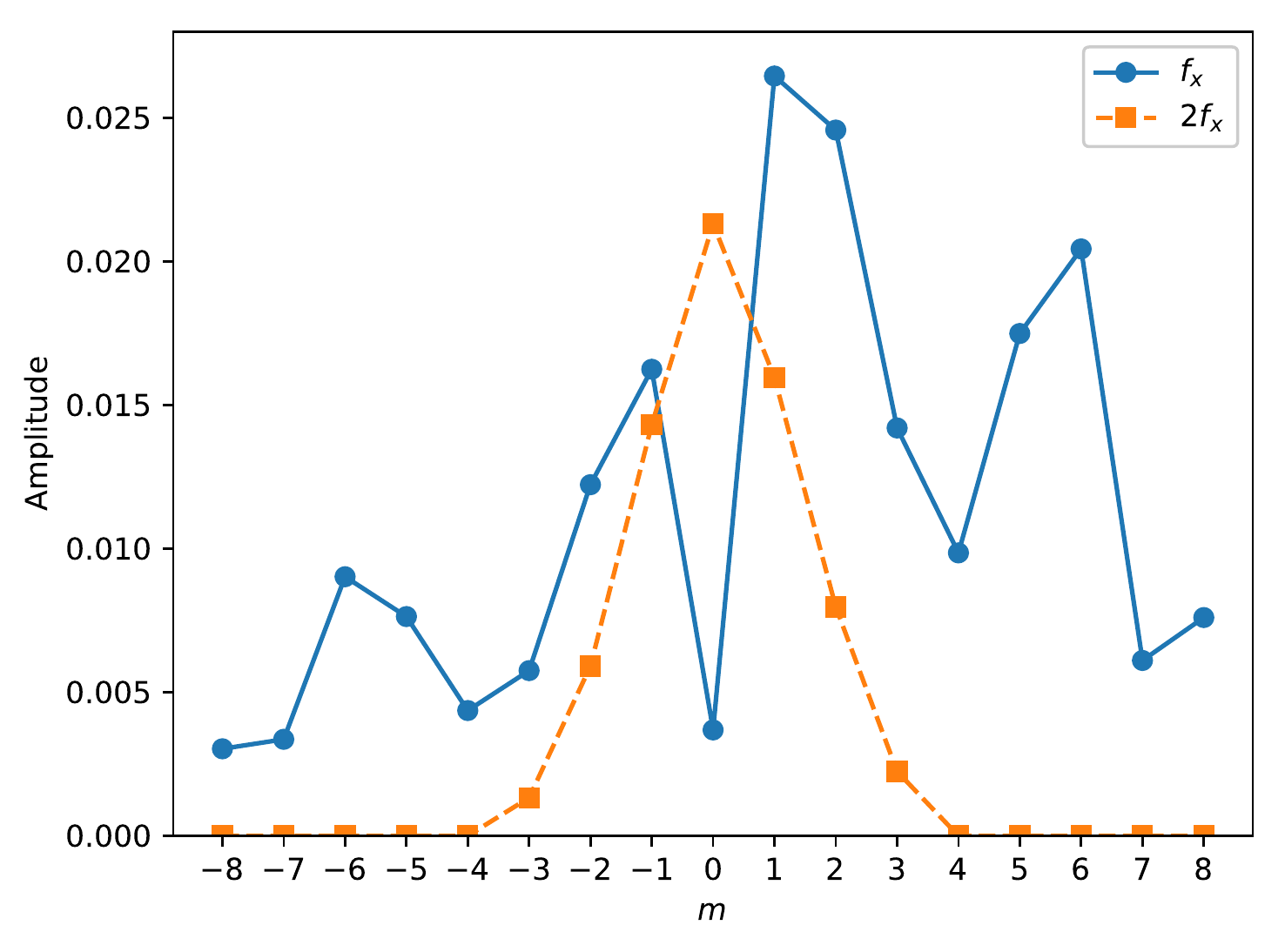}
\caption{Amplitudes of the non-radial mode with $\ell=8$ and its harmonic as a function of $m$.}
\label{fig.l8_amplitudes}
\end{figure}

\begin{figure}
\centering
\includegraphics[width=0.5\textwidth]{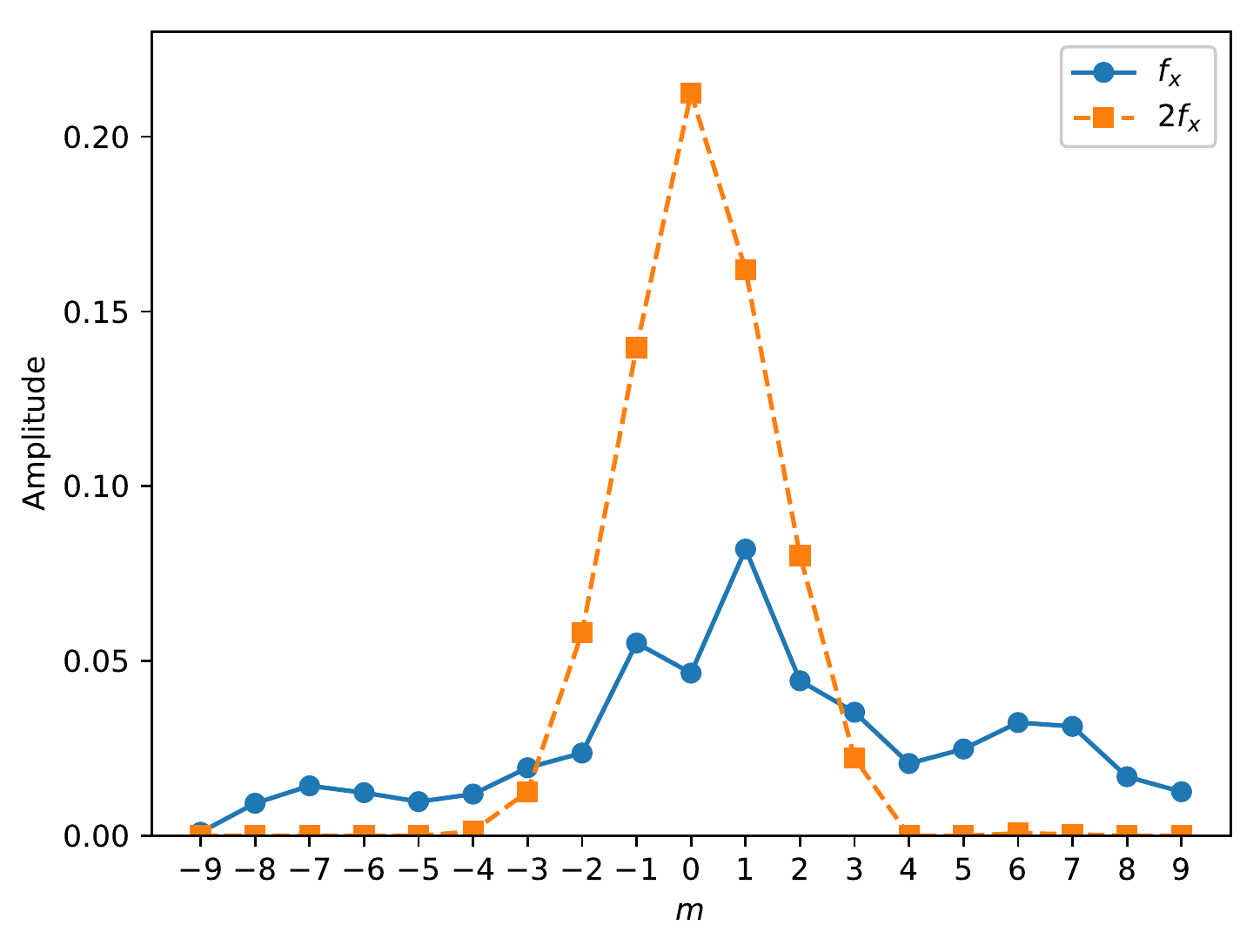}
\caption{Amplitudes of the non-radial mode with $\ell=9$ and its harmonic as a function of $m$.}
\label{fig.l9_amplitudes}
\end{figure}

\subsection{Analysis of spectra}
For the LPV analysis and the frequency search, we used the package FAMIAS as well. We analysed the mean Fourier spectrum across the line profile, the 2D spectra and performed a Fourier analysis of the first three moments, to check which approach can be more successful in detecting the additional mode. A similar comparative analysis was also carried out by \cite{depauw}, who discussed that, depending on the pulsation modes, some approaches work better than others. However, the authors did not run simulations for degrees $\ell$ as high as we are investigating during this work.

Along with the Fourier analysis, we used a consecutive prewhitening method. In analogy with the study of the photometric data, we prewhiten the data with the least-squares fit of a set of frequencies detected in the Fourier analysis. Then we performed the Fourier analysis on the residuals of the fit. This process was repeated iteratively.

\section{Results}\label{sec.Results}

In Sec.~\ref{Subsec.perfect} we show our results for the analysis of the line profile variations in the ``perfect'' noiseless case. We investigated whether is a priori possible to detect the additional modes. In Sec.~\ref{Subsec.real} we apply the same method of simulations and analysis for a set of spectra with realistic timing, including gaps in the data, and with added noise. In Sec.~\ref{Subsec.lowdeg} we show, for the selected combination of $\ell$ and $m$, the differences between signatures of non-radial modes with a lower degree and the degrees investigated in this study.

\subsection{Line profile variations in the ``perfect'' noiseless case}\label{Subsec.perfect}
In this section, we check whether it is possible to detect the additional non-radial mode signatures in case of a ``perfect'' set of spectra. 
In Fig.~\ref{fig.lpv}, we present an example of simulated line profile variations. The line profiles for pulsations in the radial first overtone mode are shown in the left panel. The  middle panel shows profiles in case of double-periodic modulation in the radial mode and non-radial mode with $\ell=7$ and $m=0$. The differences between these two cases are not easy to notice at the first glimpse. They are presented in the right panel of Fig.~\ref{fig.lpv}. 

As is visible in Fig.~\ref{fig.lpv}, the high amplitude of the dominant first-overtone could hamper the detection of the low-amplitude non-radial mode in the power spectra. For this reason, we subtracted the line profiles corresponding to pulsations in the first overtone from line profiles corresponding to double-mode pulsations before further analysis. We then performed a frequency analysis of the resulting set of line profiles. This approach significantly decreased the amplitude of the signals corresponding to variability with the radial mode and made it possible to determine whether the non-radial mode is detectable. In Fig.~\ref{fig.l7_fourier} we plot the results of the frequency analysis for a non-radial mode $\ell=7$ and all possible values of $m$, which are indicated in the top right of all panels. Note that what is shown in Fig.~\ref{fig.l7_fourier} is not the power spectrum after applying the Fourier transform, but we plotted the amplitudes and the frequencies of the subsequently detected signals in the course of our analysis. The amplitudes are mean amplitudes across the line profile because we used the mean Fourier spectra across the line profiles for this analysis. For $\ell=7$, we detect the first-overtone signal together with its harmonics. Depending on $m$, we detect the non-radial mode $f_x$ only, or together with its harmonic, 2$f_x$. We also detect combinations of the radial and non-radial mode. Positions of the first overtone, the non-radial mode and its harmonic are marked with blue, grey and red bands, respectively. Harmonics of these signals and combination frequencies are marked with dashed lines. Amplitudes of $f_x$ and 2$f_x$ change with $m$. For low $|m|$, 2$f_x$ has higher amplitude than $f_x$. The amplitude of 2$f_x$ decreases with increasing $|m|$. The signal at 2$f_x$ is marginally visible for $m=4$ (however not for $m=-4$), and for $|m|$ larger than 4 it was not detected. A similar behavior of the non-radial mode and its harmonic is visible for $\ell=8$, presented in Fig.~\ref{fig.l8_fourier} and for $\ell=9$, presented in Fig.~\ref{fig.l9_fourier}. The signal at 2$f_x$ is significantly higher than the signal at $f_x$ for $m=0$, and it is not possible to detect any signal at 2$f_x$ for $|m|$ higher than 4.

The variations in the detected amplitudes of the non-radial mode and its harmonic as a function of $m$ are presented in Fig.~\ref{fig.l7_amplitudes} for $\ell=7$, in Fig.~\ref{fig.l8_amplitudes} for $\ell=8$ and in Fig.~\ref{fig.l9_amplitudes} for $\ell=9$. The harmonic of the non-radial mode is always the highest for $m=0$, and it is not detectable for $|m|>4$. In our simulations, its amplitude is also slightly larger for positive $|m|$. The non-radial mode's amplitude is slightly higher for positive $|m|$, but its dependence on $|m|$ is very different from that of the harmonic of the non-radial mode. Interestingly, for $m=0$, the amplitude of the harmonic is the highest as mentioned above, but the amplitude of the non-radial mode is lower than for $|m|=1$. It is clear that the detection of the non-radial mode and/or its harmonic both depend on $|m|$. Only for a few combinations of $\ell$ and $m$, it would be possible to detect both signals. This feature can help in identifying the additional mode if it is detected.

\subsection{Line profile variations for lower degree modes}\label{Subsec.lowdeg}

\begin{figure}
\centering
\includegraphics[width=0.5\textwidth]{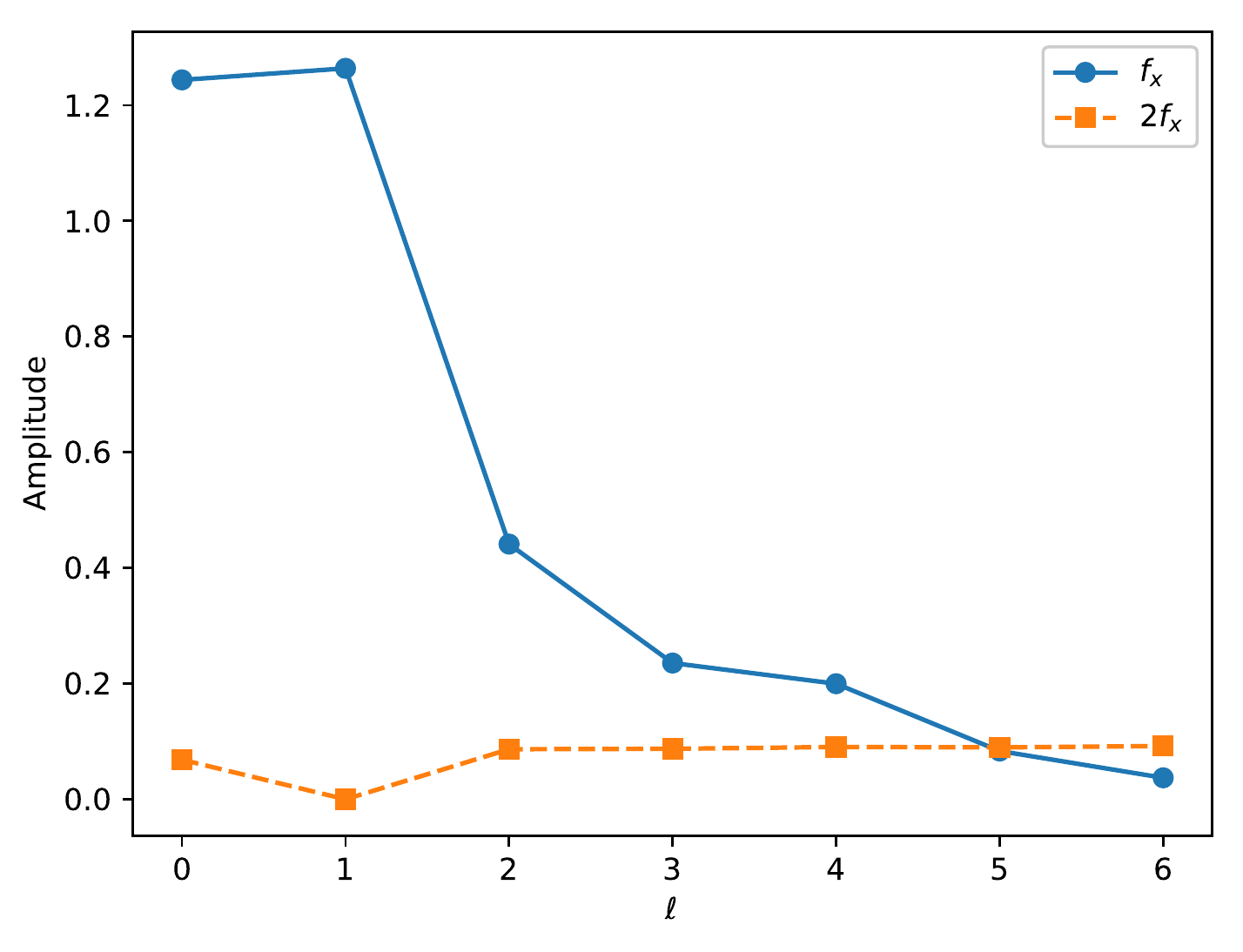}
\caption{Amplitudes of the non-radial mode and its harmonic for degrees $\ell$ from 0 to 6 and $m=0$.}
\label{fig.lowl}
\end{figure}

We investigated non-radial modes' signatures in the simulated line profile variations of degrees smaller than our non-radial modes of interest based on the model by \cite{dziembowski}. We checked $\ell$ from 0 to 6 and set $|m|=0$ for all degrees. In Fig.~\ref{fig.lowl} we plotted how the amplitudes of non-radial modes and its harmonics change with $\ell$. Up to $\ell=5$, the amplitude of the non-radial modes is higher. Still, for $\ell=6$, the amplitude of the harmonic of the non-radial mode is higher than the amplitude of the non-radial mode itself, similarly to what we observe for $\ell=7,8,9$.

\subsection{Dependence on inclination}
The Cepheid, OGLE-LMC-CEP-2532, which was used as a reference in this study, has a determined inclination of 85 degrees. In the simulations discussed above, we used an inclination of 45 degrees, but we also studied how the inclination influences the visibility of the non-radial mode and its harmonic. We simulated sets of spectra for $\ell=7,8,9$ and $|m|=0,\ell$ or $-\ell$, for inclinations set to 5, 45 or 85 degrees.

\begin{figure}
\centering
\includegraphics[width=0.5\textwidth]{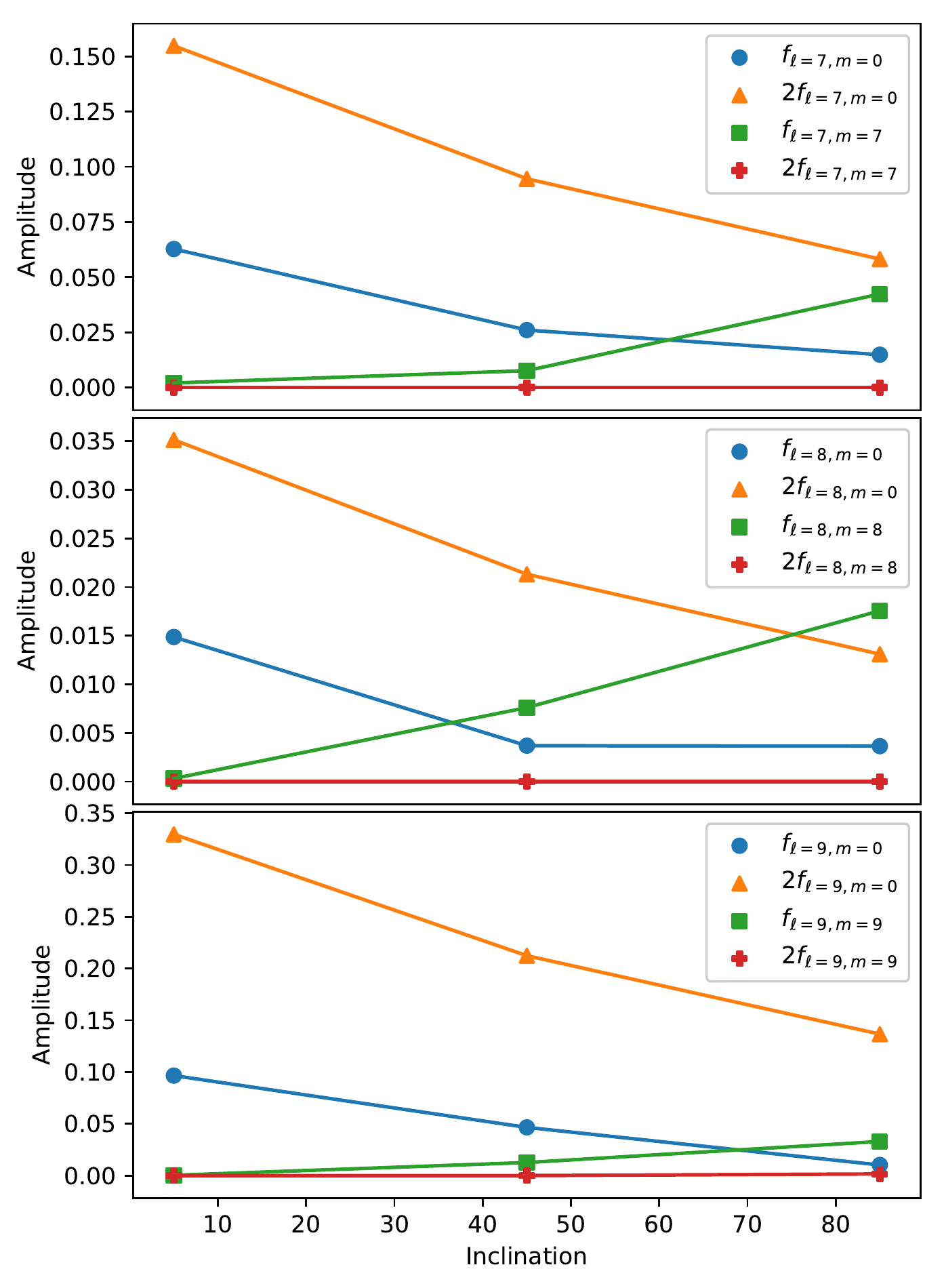}
\caption{The amplitudes of the non-radial modes and its harmonic as function of amplitudes for $\ell=7$ and $m=0$ or $m=\ell$ (top panel), $\ell=8$ and $m=0$ or $m=\ell$ (middle panel), $\ell=9$ and $m=0$ or $m=\ell$ (bottom panel).}
\label{fig.inclination}
\end{figure}

In Fig.~\ref{fig.inclination} we plot the amplitudes as a function of inclination for $\ell=7$ (top panel), $\ell=8$ (middle panel) and $\ell=9$ (bottom panel). Different line colors and types of markers were used to show results for different pairs of $m$ and $\ell$, as indicated in the key. As expected from the previous results, the signals corresponding to 2$f_{\ell=7,m=7}$, 2$f_{\ell=8,m=8}$ and 2$f_{\ell=9,m=9}$ were not detected for any inclination. The amplitudes of signals corresponding to $f_{\ell=7,m=7}$, $f_{\ell=8,m=8}$ and $f_{\ell=9,m=9}$ are too small to be detected at a low inclination of 5 degrees, but their detectability increases with increasing inclination. On the other hand, the signals corresponding to $f_{\ell=7,m=0}$, $f_{\ell=8,m=0}$ and $f_{\ell=9,m=0}$ are detectable for a low inclination of 5 degrees, and their amplitudes decrease with increasing inclination. The harmonics of these signals,i.e., signals at 2$f_{\ell=7,m=0}$, 2$f_{\ell=8,m=0}$ and 2$f_{\ell=9,m=0}$ have higher amplitudes than the non-radial modes themselves for all values of inclination and also their amplitudes decrease with increasing inclination.

Since we do not know which $m$ values are selected by stars with the additional non-radial modes, the inclination is another factor which determines whether signals can be detected or not. In the case of the representative Cepheid model studied here, an inclination of 85 degrees might be a significant factor that explains why there is no additional mode detected in this star, even though it is a first-overtone Cepheid with potentially an additional mode.

\subsection{Line profile variations in the "realistic" case}\label{Subsec.real}

From the previous sections, it is clear that even using a large number of spectra with continuous sampling it is difficult to detect signals corresponding to the suspected high-order $\ell=7,8,9$ non-radial modes and their harmonics. Therefore, detecting these signals in real spectra will be even more challenging. We studied a few sets of spectra with a more realistic sampling and with decreased signal-to-noise level to check whether it would be possible to detect signatures of these non-radial modes in real observations of first-overtone Cepheids. 

\begin{figure}
\centering
\includegraphics[width=0.5\textwidth]{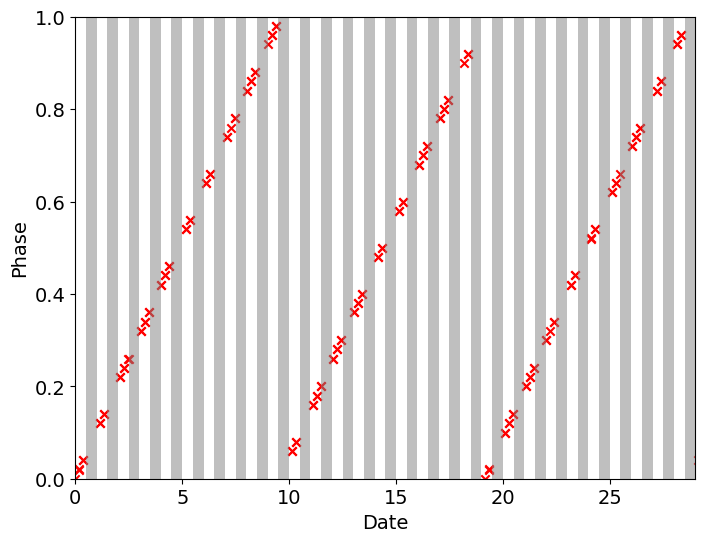}
\caption{Example sampling of our simulated spectra, which could be used as a guideline for future observations. The red crosses mark the times of 'data acquisition' or simulations. The grey areas indicate day time.}
\label{fig.times_beat_30d}
\end{figure}

First, we checked whether the harmonic of the non-radial mode, which is the most often detected signal in the photometric data, can be detected. In Tab.~\ref{tab.real} we summarize our results. The first two columns correspond to the number of nights during which the simulated observations are carried out and the number of collected spectra during the whole campaign. For each number of nights and spectra, we changed the signal-to-noise level. Without explicitly including noise in the simulated set of spectra, the mean signal-to-noise level is around 430. We also checked signal-to-noise levels of 300, 200, 150 and 100. The first five rows of Tab.~\ref{tab.real} correspond to observations each night for 30 or 50 nights. The sampling of spectra was chosen to cover all phases of the beating period between the first overtone mode and the non-radial mode. Fig.~\ref{fig.times_beat_30d} shows how phases of the beating period are covered with observations during 30 days. Grey areas indicate day gaps. In Tab.~\ref{tab.real} we indicated whether the non-radial mode was detected with 'yes' or 'no' using the mean Fourier spectrum across the line profile. A '{\bf yes}' with bold font indicates that it was possible to detect the non-radial mode only in the Fourier spectrum of the second moment. With more than 200 spectra collected during 50 days it is possible to detect the additional mode, even for a signal-to-noise level of 150, but for all these cases discussed above it is not possible to detect the mode for a  signal-to-noise level of 100.

\begin{table}
 \centering
  \caption{Detection of harmonic of the non-radial mode for simulations for $\ell=7$ and $m=0$ for real sampling of the data. The table indicates for which signal to noise ratio (S/N) and number of nights of observations the harmonic was detected. Normal font 'yes' means detection in the mean Fourier spectrum. Bold font '{\bf yes}' means the mode was detected in the Fourier spectrum of the second moment.}
  \label{tab.real}
  \centering
  \begin{tabular}{@{}ll|ccccc@{}}
  \hline
&  &\multicolumn{5}{c}{S/N}\\
N\textsubscript{nights} & N\textsubscript{spectra} & $\sim$430 & 300 & 200 & 150 & 100\\ 
\hline
30 & 147 & yes & {\bf yes} & {\bf yes} & no & \\
50 & 74 & yes & no & &&\\
50 & 96 & yes & {\bf yes} & no &&\\
50 & 121 & yes & {\bf yes} & {\bf yes} & no &\\
50 & 244 & yes & yes & yes & {\bf yes} & no\\
40 of 60 & 200 & yes & {\bf yes} & no &  & \\
10 of 60 & 100 & no &  &  &  & \\
40 of 60 & 400 & yes & yes & yes & yes & {\bf yes} \\
 \hline

\end{tabular}
\end{table}

\begin{figure}
\centering
\includegraphics[width=0.5\textwidth]{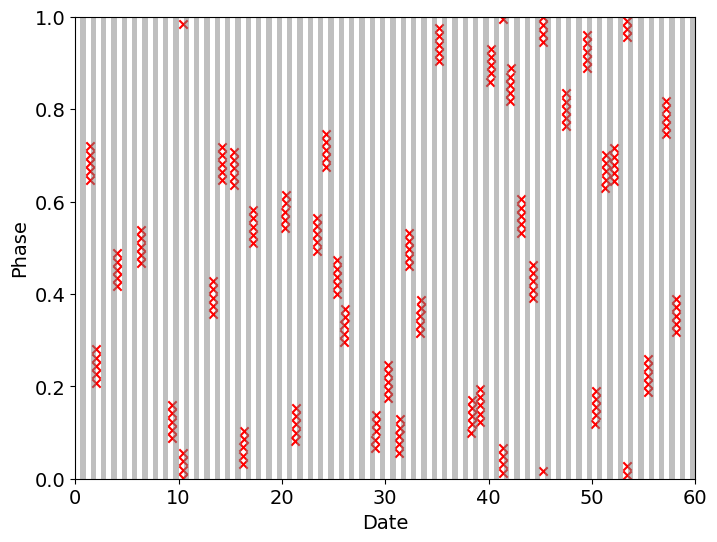}
\caption{Example sampling of our simulated spectra. The red crosses mark the times of 'data acquisition' or simulations. The grey areas indicate day time. Forty observing nights were randomly selected out of a 60 night period. Five spectra were taken during each observing night. We required that the spectra cover the whole beating period.}
\label{fig.times_random40_over60}
\end{figure}

The last three rows of Tab.~\ref{tab.real} shows cases of campaigns where the observations were not carried out every night, which simulates realistic weather conditions from most Earth-bound observing sites. In these simulations, 10 or 40 nights were randomly selected from the 60 day period spanning the observations. The resulting number of spectra was 100, 200 or 400. In Fig.~\ref{fig.times_random40_over60} we present an example for an observational campaign of 40 nights randomly selected from a 60 night period with five spectra collected during each observing night. Again, the condition was to cover the whole beating period. As visible in Tab.~\ref{tab.real}, it is significantly harder to detect the additional mode with gaps in the data, because even for 200 spectra, the mode was detected for signal-to-noise of 300 and not for lower.

Secondly, we checked the possibility to detect the non-radial mode itself for different numbers of observing nights and different numbers of collected spectra. We also  adopted the observing times from \cite{OGLE-LMC-CEP-2532} in our simulations. We performed the frequency search using the analysis of the first moment because this method was the most effective in detecting the non-radial mode. We also checked how the possibility to detect the mode changes with the azimuthal order of the mode. Our results are presented in Tab.~\ref{tab.real7} for the mode with $\ell=7$, in Tab.~\ref{tab.real8} for the mode with $\ell=8$, and in Tab.~\ref{tab.real9} for the mode with $\ell=9$. The detection of the non-radial mode was not possible in the data with added noise in the case of the azimuthal order $m=0$ for all considered degrees. In the case of the mode $\ell=7$, the detection was possible for azimuthal order $m$= --1, 1, 2, and 4. The amplitude of the signal for azimuthal order $m=3$ is lower than in the case of the azimuthal order $m=4$ (see Fig.~\ref{fig.l7_amplitudes}). Indeed, it was not possible to detect the mode of azimuthal order $m=3$ in the data with added noise. The detection is the easiest for an azimuthal order $m=1$. For a total of 244 spectra the detection was possible even for relatively low $S/N=100$ and for observing times from \cite{OGLE-LMC-CEP-2532} for $S/N=300$.

For the non-radial mode of degree $\ell=8$, the detection in the data with added noise was again not possible for azimuthal order $m=0$. For an azimuthal order $m=1$, the detection was possible down to $S/N=150$ for a total of 244 spectra taken every night and for 200 spectra taken during 40 nights selected from a 60-night period. Also for the azimuthal order $m=1$ the detection was possible in the case of sampling from \cite{OGLE-LMC-CEP-2532} with $S/N=300$.

In the case of the non-radial mode of degree $\ell=9$, the observations are similar to modes of degrees 7 and 8. Detection of the mode was possible in the spectra with added noise. In the case of the azimuthal order $m=1$, it was possible for $S/N=200$ for a total of 244 and 200 spectra. It was still possible to detect the mode for the time sampling from \cite{OGLE-LMC-CEP-2532} for spectra with $S/N=300$.

\begin{table}
 \centering
  \caption{Detection of the non-radial mode for simulations for $\ell=7$ for the real sampling of the data. The first three columns provide the azimuthal order, {\it m}, the number of nights, and the total number of spectra. The next columns indicate for which signal to noise ratio (S/N) and the number of nights of observations for which the harmonic was detected. The frequency search was performed using the first moment.}
  \label{tab.real7}
  \centering
  \begin{tabular}{@{}lll|ccccc@{}}
  \hline
& &  &\multicolumn{5}{c}{S/N}\\
$m$ & N\textsubscript{nights} & N\textsubscript{spectra} & no noise & 300 & 200 & 150 & 100\\
\hline
0 &  50 & 244 & yes & no & --  &  -- & -- \\
1 &  50 & 244 & yes & yes & yes & yes & yes\\
-1 &  50 & 244 & yes & yes & yes & yes & no \\
2 &  50 & 244 & yes & yes & yes & no & -- \\
3 &  50 & 244 & yes & no & --  &  -- & -- \\
4 &  50 & 244 & yes & yes & yes & no &  --\\
0 &  \multicolumn{2}{l}{Pilecki et al. (2015)} & yes & no & -- & -- &--  \\
1 &  \multicolumn{2}{l}{Pilecki et al. (2015)} & yes & yes & no & -- & -- \\
 \hline

\end{tabular}
\end{table}

\begin{table}
 \centering
  \caption{The same as Tab.~\ref{tab.real7}, but for $\ell=8$.}
  \label{tab.real8}
  \centering
  \begin{tabular}{@{}lll|ccccc@{}}
  \hline
& &  &\multicolumn{5}{c}{S/N}\\
$m$ & N\textsubscript{nights} & N\textsubscript{spectra} & no noise & 300 & 200 & 150 & 100\\
\hline
0 & 50 & 244 & yes & no & -- & --  & -- \\
1 & 50 & 244 & yes & yes & yes & yes & no\\
2 & 50 & 244 & yes & yes & no & --  & -- \\
3 & 50 & 244 & yes & no & -- & -- & -- \\
0 &  50 & 50 & yes & no &  -- & --  & -- \\
1 &  50 & 50 & yes & yes & yes & no &  -- \\
1 &  40 z 60 & 200 & yes & yes & yes & yes & no\\
2 &  40 z 60 & 200 & yes & no & -- & -- & -- \\
0 &  \multicolumn{2}{l}{Pilecki i in. (2015)} & no & --  & -- &  -- &  -- \\
1 &  \multicolumn{2}{l}{Pilecki i in. (2015)} & yes & yes & no & --  & -- \\
 \hline

\end{tabular}
\end{table}

\begin{table}
 \centering
  \caption{The same as Tab.~\ref{tab.real7}, but for $\ell=9$.}
  \label{tab.real9}
  \centering
  \begin{tabular}{@{}lll|ccccc@{}}
  \hline
& &  &\multicolumn{5}{c}{S/N}\\
$m$ & N\textsubscript{nights} & N\textsubscript{spectra} & no noise & 300 & 200 & 150 & 100\\
\hline
0 & 50 & 244 & yes & no & -- & -- & -- \\
1 & 50 & 244 & yes & yes & yes & no & -- \\
1 &  50 & 50 & yes & yes & no & --  & -- \\
0 & 40 z 60 & 200 & yes & no & -- & -- & -- \\
1 &  40 z 60 & 200 & yes & yes & yes & no & -- \\
2 &  40 z 60 & 200 & yes & no & -- & -- & -- \\
4 &  40 z 60 & 200 &  yes  & no  & --  & --  & -- \\
1 & \multicolumn{2}{l}{Pilecki i in. (2015)} & yes & yes & no & -- &  -- \\
 \hline

\end{tabular}
\end{table}

\section{Discussion}\label{sec.discussion}

The mysterious additional signals detected in a fraction of first-overtone classical Cepheids and RR Lyrae stars form a characteristic period ratio from 0.60 to 0.65 with the main pulsation mode.  In a model proposed by \cite{dziembowski}, these detected frequencies the harmonics of non-radial modes of degrees $\ell$ 7, 8 or 9, explaining the three sequencies in the Petersen diagram \citep[see fig. 3 in][]{dziembowski}. Note that in RR Lyrae stars we also observe three similar sequences, but according to the model by \cite{dziembowski}, the top and bottom sequences correspond to degrees $\ell$ 8 and 9, whereas the middle sequence would be due to combinations between the two modes. These additional signals were first detected in photometric data of classical Cepheids and RR Lyrae stars. In this paper, we studied whether it would be possible to detect these low-amplitude high-degree non-radial modes in spectroscopy. If this is true, then spectroscopic mode identification can shed some light on the nature of the additional modes. 

Our simulations and the subsequent analysis indicated that the visibility of the mode and its harmonic for degrees $\ell=7,8,9$ depends strongly on the azimuthal number $m$ of the mode. We do not know what the exact configuration in the star is, but for this study, we assumed only one excited additional mode with selected $\ell$ and $|m|$. We observed that for different combinations of the degree $\ell$ and the azimuthal order $|m|$, it might be possible to detect the non-radial mode only, its harmonic only or both signals at the same time. In fact, this is also observed in the photometric study of classical Cepheids and RR Lyrae stars. In most cases, we observe the harmonic, in some stars, we observe harmonic and the non-radial mode with the amplitude of the non-radial mode sometimes higher than the amplitude of its harmonic. Also, several RR Lyrae stars are thought to be stars where only the non-radial mode is detected without its harmonic \citep[for a discussion of this in RR Lyrae stars see][]{netzel_census}.

In our study we took OGLE-LMC-CEP-2532, a known first-overtone Cepheid in a binary system as a reference because if its well-determined physical parameters. The star was studied spectroscopically by \cite{OGLE-LMC-CEP-2532}. In order to determine the system parameters, the authors used 49 high-resolution spectra. There are 1322 ground-based photometric measurements in the $I$ filter and 231 in the $V$ filter available for OGLE-LMC-CEP-2532. Since we used this star as a reference, we checked if the additional signal is detectable in photometric data and concluded it is not. However, we assume that results from ground-based photometry, where a high fraction of stars show the additional mode, and space-based photometry, where almost all stars show it, suggest that the mode is present in the majority of the first-overtone stars. In the case of ground-based photometry, the high noise and the daily aliases in the power spectra hamper the detection of the additional mode, and for this star this might also be the case. On the other hand, the number of data points available for this star is high enough to expect a detectable signal, if it is there. We therefore simulated a set of spectra for the times of observation used in \cite{OGLE-LMC-CEP-2532}. 
With such a small number of spectra, it is impossible to detect the additional signal (harmonic of the additional non-radial mode) even without including the noise in simulations. It was possible to detect the non-radial mode itself, but only for azimuthal order $m=1$ and for relatively high $S/N$.


The inclination is another factor which affects the detectability of the non-radial mode. In the case of OGLE-LMC-CEP-2532, the inclination is 85 degrees, which based on this study significantly hampers the detection of the non-radial mode. {This may also explain why we do not observe the additional frequency in the photometric data of the star. }

Investigating sets of spectra with simulated gaps in the data and with included noise shows that detecting this mode in real data is an extremely challenging task and depends strongly on the azimuthal order. Only a large number of spectra would allow detection of these additional signals. If the data set has large gaps, detection would require a set of 200 high-resolution spectra. Also, the signal-to-noise level has to be high, well above 200 and preferably 300 for a reliable detection. The task is challenging but not impossible.  Obtaining such a high signal-to-noise ratio is difficult in a single spectrum, but the least-squares deconvolution (LSD) method can be of help. \cite{lsd} tested the improved version of the method and obtained 5-15 times gain for the signal-to-noise ratio in the LSD spectrum. This tool could easily help to obtain the signal-to-noise ratio required to detect the additional mode.

During simulations, we assumed pulsations in two modes: in the radial mode and in the non-radial mode of a certain degree and azimuthal order. However, in the star, modes of different azimuthal orders might be present, interacting with each other. The prediction of such an interaction between $2\ell+1$ modes is not trivial and was neglected in the presented analysis.

Our study was focused on classical Cepheids, but these additional signals are also present in RR Lyrae stars as was mentioned before. We focused on classical Cepheids because we used the estimation for the velocity amplitude of the non-radial mode applicable to Cepheids \citep[see Eg. 8 in][]{dziembowski2012}. However, the results should be similar in case of an RR Lyrae star model for non-radial modes with $\ell=8,9$. Classical Cepheids, due to their brightness and longer periods, are more attractive targets for an observing campaign focused on the nature of the additional modes.

\section{Summary}\label{sec.summary}

In photometric data of first-overtone classical Cepheids and RR Lyrae stars there are detected additional signals which are suggested to be due to pulsations in non-radial modes of degrees $\ell=7,8,9$ (classical Cepheids) or $\ell=8,9$ (RR Lyrae stars) \citep{dziembowski}. The main goal of our study was to determine whether these additional modes can be detected and possibly identified in spectroscopic data.

We studied simulated sets of spectroscopic line profiles (LPV) for classical Cepheid. The parameters for our simulations were based on the data for the first-overtone classical Cepheid OGLE-LMC-CEP-2532 \citep{OGLE-LMC-CEP-2532}, and an estimation for the velocity amplitude of the non-radial modes was derived from \cite{dziembowski2012}. We studied sets of spectra with dense, almost continuous, sampling, and without noise - so-called 'perfect' cases. We also studied how a decreasing signal-to-noise level and sampling with gaps affect the detectability of the mode and its harmonic. Our findings can be summarized as follows:

\begin{itemize}
\item In the power spectra derived from the LPV we detected, besides the expected first overtone mode and the additional non-radial mode, also the harmonics of both modes and combination signals.
\item Depending on the azimuthal order $m$, the amplitudes of the non-radial mode and its harmonic change. For $|m|>4$ the harmonic is not detected and it has the highest amplitude for $m=0$ for all degrees $\ell$.  The amplitude of the non-radial mode is lower for $m=0$ than for $|m|=1$ and varies depending on $|m|$.
\item To investigate the spectroscopic peculiarities of the higher-degree modes suggested in the model by \cite{dziembowski}, we also studied other degrees of modes $\ell$ from 0 to 6, and azimuthal number $m=0$, and observed that for modes with $\ell$ higher than 5, the harmonic of the non-radial mode has a higher amplitude than the mode.
\item The inclination under which the non-radial modes are observed is an important factor determining whether it is possible to detect the non-radial mode or its harmonic. The amplitude of the non-radial mode and its harmonic decreases for $m=0$ with increasing inclination. The amplitude of the non-radial mode for $m=\ell$ increases with increasing inclination. 
\item A small number of spectra (around 50) is not enough to detect the harmonic of the non-radial mode. Also, a higher amount of spectra (around 200) with random sampling are not enough to detect this signal. The observing campaign must be tailored to cover the whole beating period of the first overtone mode and the non-radial mode. Even in this case, the number of obtained spectra must be large: if observations can be carried out every night, the required number of spectra can be below 100; if there are longer gaps between observing nights, the number of spectra must be above 100, preferably around 200. 
\item The azimuthal order is an important factor determining the possibility to detect the non-radial mode. It is easiest to detect the mode for $m=1$ for all considered degrees. For the value of the azimuthal order other than $m=1$, the total number of spectra and signal-to-noise ratio of the spectra typically have to be high. Around 200 spectra and $S/N=300$ are preferred.
\item The signal-to-noise level of the spectra has to be high, well above 100. In the case of a smaller number of spectra, the signal-to-noise ratio should be around 300. Such a high signal-to-noise ratio can be obtained using the least-squares deconvolution method.
\end{itemize}

In our quest to better understand the additional frequencies detected in first-overtone classical Cepheids and RR Lyrae stars, spectroscopic time series offer an unexplored and promising way forward.
The results of our extensive simulations of the line profile variations (LPV) of a first-overtone Cepheid pulsating with an additional low-amplitude non-radial mode can be used to plan a spectroscopic observing campaign tailored to uncover the nature of these mysterious additional modes. 
Moreover, our findings indicate that the detected amplitude ratios for the suspected non-radial mode and its harmonic can yield some constraints on the degree $\ell$ and the azimuthal order $m$ of the mode.
It is clear that a spectroscopic identification of the mysterious modes in first-overtone Cepheids and RR Lyrae stars will not be an easy task, but not entirely impossible especially given the facilities and techniques available nowadays.

\section*{Acknowledgements}
HN is supported in by the Polish Ministry of Science and Higher Education under grant 0192/DIA/2016/45 within the Diamond Grant Programme for years 2016-2021 and Foundation for Polish Science (FNP). We also gratefully acknowledge the international exchange project VS.091.16N entitled "Probing Cosmic Standard Candles in the Space Age"  between the Belgian-Flemish Fund for Scientific Research (FWO) and the Polish Academy of Sciences (PAS) which enabled the collaborations leading to this paper. This project has been supported by the Lend\"ulet Program of the Hungarian Academy of Sciences, project No. LP2018-7/2020.

\section*{Data availability}
This paper uses the publicly available code to generate simulated spectroscopic series. Input data for the code are explicitly provided inside the paper.




\bibliographystyle{mnras}
\bibliography{biblio} 




\appendix

\section{Supplementary figures}

\begin{figure*}
\begin{minipage}{180mm}
\centering
\includegraphics[width=\textwidth]{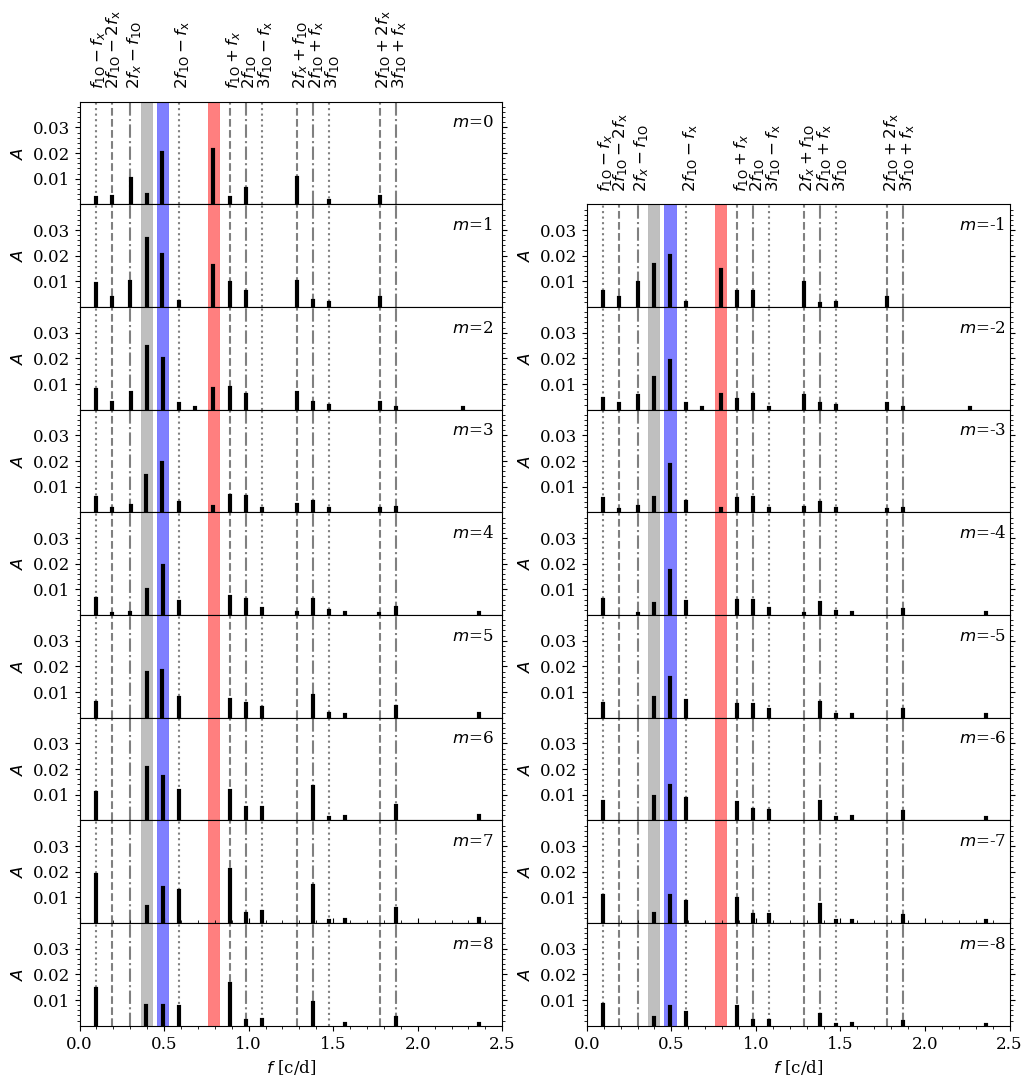}
\caption{The same as for Fig.~\ref{fig.l7_fourier}, but for $\ell=8$.}
\label{fig.l8_fourier}
\end{minipage}
\end{figure*}

\begin{figure*}
\begin{minipage}{180mm}
\centering
\includegraphics[width=\textwidth]{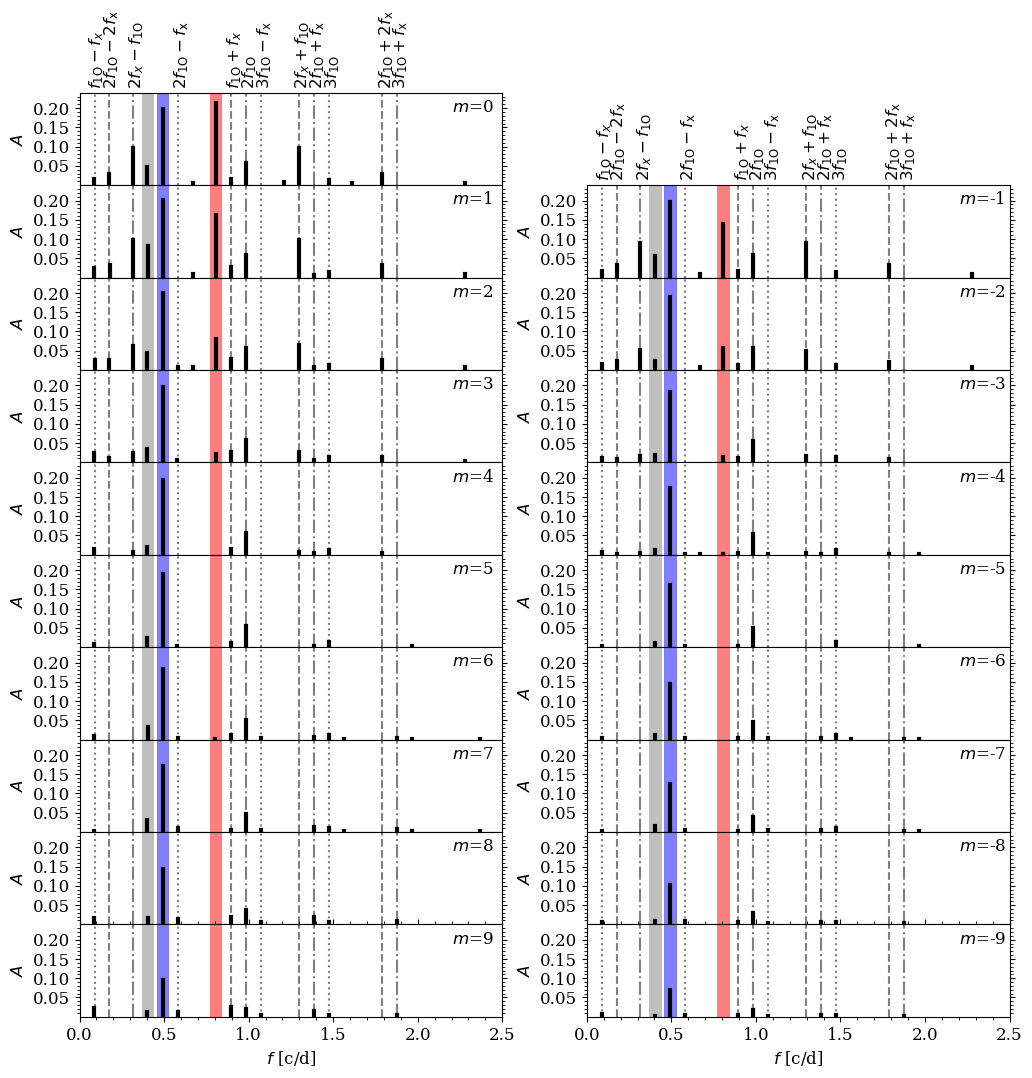}
\caption{The same as for Fig.~\ref{fig.l7_fourier}, but for $\ell=9$.}
\label{fig.l9_fourier}
\end{minipage}
\end{figure*}


\bsp	
\label{lastpage}
\end{document}